\documentclass[aps,pra,twocolumn,showpacs,superscriptaddress,floatfix]{revtex4-1}
\usepackage{graphicx,amsmath,amssymb}
\usepackage[usenames]{color}
\usepackage[dvipsnames]{xcolor}
\usepackage[unicode=true,pdfusetitle,bookmarks=false,breaklinks=false,pdfborder={0 0 1},backref=false,colorlinks,urlcolor=Black, linkcolor=blue,citecolor=blue]{hyperref}
\begin{document}

\title{Robust topological feature against non-Hermiticity in Jaynes-Cummings Model
}
\author{Zu-Jian Ying }
\email{yingzj@lzu.edu.cn}
\affiliation{School of Physical Science and Technology, Lanzhou University, Lanzhou 730000, China}
\affiliation{Key Laboratory for Quantum Theory and Applications of MoE, Lanzhou Center for Theoretical Physics, and Key Laboratory of Theoretical Physics of Gansu Province, Lanzhou University, Lanzhou, Gansu 730000, China}

\begin{abstract}
The Jaynes-Cummings Model (JCM) is a fundamental model and building block for light-matter interactions, quantum information
and quantum computation. We analytically analyze the topological feature manifested by the JCM in the presence of non-Hermiticity which may be effectively induced by dissipation and decay rates. Indeed, the eigenstates of the JCM are topologically characterized by spin windings in two-dimensional plane. The non-Hermiticity tilts the spin winding plane and induces out-of-plane component, while the topological feature is maintained. In particular, besides the invariant spin texture nodes, we find a non-Hermiticity-induced reversal transition of the tilting angle and spin winding direction with a fractional phase gain at gap closing, a partially level-independent reversal transition without gap closing, and a completely level-independent super invariant point with untilted angle and also without gap closing. Our result demonstrates that the topological feature is robust against non-Hermiticity, which would be favorable in practical applications. On the other hand, one may conversely make use of the disadvantageous dissipation and decay rates to reverse the spin winding direction, which might add a control way for topological manipulation of quantum systems in light-matter interactions.
\end{abstract}
\pacs{ }
\maketitle


\section{Introduction}

The continuous theoretical progresses~\cite%
{Braak2011,Solano2011,Boite2020,Liu2021AQT} and experimental advances~\cite%
{Ciuti2005EarlyUSC,Aji2009EarlyUSC,Diaz2019RevModPhy,Kockum2019NRP,Wallraff2004,Gunter2009, Niemczyk2010,Peropadre2010,FornDiaz2017,Forn-Diaz2010,Scalari2012,Xiang2013,Yoshihara2017NatPhys,Kockum2017,Bayer2017DeepStrong}
over the last two decades have made the systems in light-matter interactions
an ideal platform for simulations of traditional states of matter and
explorations of novel quantum states and quantum technologies. Indeed,
both
few-body quantum phase transitions~\cite{
Liu2021AQT,Ashhab2013,Ying2015,Hwang2015PRL,Ying2020-nonlinear-bias,Ying-2021-AQT,LiuM2017PRL,Hwang2016PRL,Irish2017, Ying-gapped-top,Ying-Stark-top,Ying-Spin-Winding,Ying-2018-arxiv,Ying-Spin-Winding,Ying-JC-winding,Grimaudo2022q2QPT,Grimaudo2023-Entropy}
and topological phase transitions~\cite%
{Ying-2021-AQT,Ying-gapped-top,Ying-Stark-top,Ying-Spin-Winding,Ying-JC-winding}
have been found in light-matter interactions and applications for critical
quantum metrology~\cite%
{Garbe2020,Garbe2021-Metrology,Ilias2022-Metrology,Ying2022-Metrology} have
been proposed, with the advantage of high controllability and tunability.

The most fundamental models of light-matter interactions are the quantum Rabi
model (QRM)~\cite{rabi1936,Rabi-Braak,Eckle-Book-Models} and the
Jaynes-Cummings model (JCM)~\cite{JC-model,JC-Larson2021}, they are also
fundamental building blocks for quantum information and quantum computation.~%
\cite{Diaz2019RevModPhy,Romero2012,Stassi2020QuComput,Stassi2018,Macri2018}
Theoretically the milestone work~\cite{Braak2011} revealing the
integrability of the QRM has induced a massive dialogue~\cite{
Braak2011,Solano2011,Boite2020,Liu2021AQT,Diaz2019RevModPhy,Kockum2019NRP,Rabi-Braak,Braak2019Symmetry,
Wolf2012,FelicettiPRL2020,Felicetti2018-mixed-TPP-SPP,Felicetti2015-TwoPhotonProcess,Simone2018,Alushi2023PRX,
Irish2014,Irish2017,Irish-class-quan-corresp,
PRX-Xie-Anistropy,Batchelor2015,XieQ-2017JPA,
Hwang2015PRL,Bera2014Polaron,Hwang2016PRL,Ying2015,LiuM2017PRL,Ying-2018-arxiv,Ying-2021-AQT,Ying-gapped-top,Ying-Stark-top,Ying-Spin-Winding,Grimaudo2022q2QPT,Grimaudo2023-Entropy,
CongLei2017,CongLei2019,Ying2020-nonlinear-bias,LiuGang2023,ChenQH2012,Zhu-PRL-2020,
e-collpase-Garbe-2017,e-collpase-Duan-2016,Garbe2020,Rico2020,
Garbe2021-Metrology,Ilias2022-Metrology,Ying2022-Metrology,
Boite2016-Photon-Blockade,Ridolfo2012-Photon-Blockade,Li2020conical,
Ma2020Nonlinear,
Padilla2022,ZhengHang2017,Yan2023-AQT,Zheng2017,Chen-2021-NC,Lu-2018-1,Gao2021,PengJie2019,Liu2015,Ashhab2013, ChenGang2012,FengMang2013,Eckle-2017JPA,Maciejewski-Stark,Xie2019-Stark,Casanova2018npj,HiddenSymMangazeev2021,HiddenSymLi2021,HiddenSymBustos2021,
JC-Larson2021,Stark-Cong2020,Cong2022Peter,Stark-Grimsmo2013,Stark-Grimsmo2014,Downing2014}
between mathematics and physics in light-matter interactions.~\cite{Solano2011} An intriguing
phenomenon in such few-body systems is the existence\cite%
{Liu2021AQT,Ashhab2013,Ying2015,Hwang2015PRL,Ying2020-nonlinear-bias,Ying-2021-AQT,LiuM2017PRL,Hwang2016PRL,Irish2017,Ying-gapped-top,Ying-Stark-top,Ying-2018-arxiv,Ying-Spin-Winding,Grimaudo2022q2QPT}
of a quantum phase transition (QPT)\cite{Sachdev-QPT} which traditionally
lies in condensed matter, despite that it might be a matter of taste to term
the transition quantum or not by considering the negligible quantum
fluctuations in the photon vacuum state.\cite{Irish2017} Here in other fields, finite-size phase transitions can occur with some level crossing, e.g. in pairing-depairing models~\cite{Ying2008PRL,Ying2007confined,Ying2006SCFM} and coupled fermion-boson models.~\cite{Stojanovic2020PRB,Stojanovic2020PRL,Stojanovic2021PRA} In light-matter interactions, the QPT occurs
in the low-frequency limit which is a replacement of thermodynamical limit
in many-body systems. It turns out that this few-body QPT can be indeed
bridged to the thermodynamical limit by the universality of the critical
exponents.\cite{LiuM2017PRL}

Interestingly both the two essentially different classes of phase
transitions can occur in systems in light-matter interactions despite their
contrary symmetry requirements. Traditional phase transitions are classified
by Landau theory which made a break through to realize that phase
transitions are accompanied with some symmetry breaking,\cite{Landau1937}
while the other class of transitions are the topological phase transitions
(TPTs)~\cite%
{Topo-KT-transition,Topo-KT-NoSymBreak,Topo-Haldane-1,Topo-Haldane-2,Topo-Wen,ColloqTopoWen2010}
which involve no symmetry breaking. The afore-mentioned QPT in the
low-frequency limit actually belongs to the Landau class, with a hidden
symmetry breaking.~\cite{Ying-2021-AQT} Phase transitions with more
varieties of symmetry-breaking patterns can also occur in the presence of
the bias and the non-linear coupling.~\cite{Ying2020-nonlinear-bias} When
the frequency is tuned to finite regime the traditional critical
universality collapses. In such a situation, the properties are diversified and apart from the
traditional QPT emerge a series of novel phase transitions, as in the
anisotropic QRM, which are found to be TPTs with the preserved parity
symmetry.~\cite%
{Ying-2021-AQT,Ying-gapped-top,Ying-Stark-top,Ying-Spin-Winding} The
topological universality in the emerging topological phases holds not only
in the anisotropy but also in the nonlinear Stark coupling.\cite%
{Ying-Stark-top} Most TPTs in these light-matter-interaction systems are
conventional ones~\cite{Ying-2021-AQT} which occur with gap closing as those
in condensed matter~\cite%
{Topo-Wen,Hasan2010-RMP-topo,Yu2010ScienceHall,Chen2019GapClosing,TopCriterion,Top-Guan,TopNori}%
. Unconventional TPTs without gap closing are also revealed~\cite%
{Ying-gapped-top,Ying-Stark-top} analogously to the unconventional cases in
the quantum spin Hall effect with strong electron-electron interactions~\cite%
{Amaricci-2015-no-gap-closing} and the quantum anomalous Hall effect with
disorder.~\cite{Xie-QAH-2021} Here the special point for these
light-matter-interaction systems is that a same system can have both the
symmetry-breaking Landau class of transitions and the symmetry-protected
TPTs by tuning the frequency,~\cite%
{Ying-2021-AQT,Ying-Stark-top,Ying-JC-winding} while conventionally they are
incompatible due to the contrary symmetry requirements. Furthermore, these
two contrary classes of transitions can even occur simultaneously at a same
finite frequency in the JCM.~\cite{Ying-JC-winding}

The topological universality in the light-matter interactions is represented
by the topological structure of the eigen wave function which universally
has a same node number at all points in a topological phase.~\cite%
{Ying-2021-AQT,Ying-gapped-top,Ying-Stark-top} As a topological quantum
number the node number also has a physical correspondence to the spin
winding~\cite{Ying-Spin-Winding,Ying-JC-winding} which is more explicit topological character. In condensed matter it has
been known that the topological properties are robust under the protection
of symmetry against perturbations, impurities and boundary conditions in
realistic systems.~\cite%
{ColloqTopoWen2010,RobustTopo-Dusuel2011,RobustTopo-Balabanov2017,RobustTopo-Multer2021}
In the quantum systems of light-matter interactions, a main concerned
aspect in practical systems is the effect of the dissipation and decay
rates.~\cite%
{Kockum2019NRP,Dissipation-Beaudoin2011,NonHermitianJCM-2022SciChina} The
dissipation and decay rates effectively induce non-Hermiticity.~\cite%
{NonHermitianJCM-2022SciChina} Nowadays non-Hermitian physics has been
attracting more and more attention.~\cite{NonHermitian-Bender2007,NonHermitian-Bender2007,NonHermitian-Review-Ashida2020} In
this situation, one may wonder how the topological feature in the
light-matter interactions is influenced by such a non-Hermiticity.

In this work, we present a study for the topological feature of the systems
in light-matter interactions in a general non-Hermiticity which may arise from dissipation of coupling and the decay rates of the qubit and the
bosonic field.~\cite{NonHermitianJCM-2022SciChina} We choose the fundamental
JCM~\cite{JC-model,JC-Larson2021} for such an investigation, considering
that the QRM has no TPTs in the ground state,~\cite%
{Ying-2021-AQT,Ying-Spin-Winding} due to the constraint of the extended~\cite%
{Ying-gapped-top} no-node theorem,~\cite{Ref-No-node-theorem} and the
non-Hermitian JCM already has the base of the experimental implementation.~%
\cite{NonHermitianJCM-2022SciChina} Since the JCM is analytically solvable,
we can see the effect of the non-Hermiticity explicitly and exactly. We find
that the non-Hermiticity only tilts the spin winding plane, while the spin
winding number is maintained for all eigen states. Besides this main trend,
we also have some other findings including invariant spin texture nodes, a
non-Hermiticity-induced reversal transition of the tilting angle and spin
winding direction at gap closing, a partially level-independent reversal
transition unconventionally without gap closing, and a completely level-independent
super invariant point with untilted angle and also without gap closing. Our
result demonstrates that the topological feature is robust against
non-Hermiticity, which would be favorable in practical applications. Also,
the found reversal transitions might add a way to quantum control by making use
of the dissipation and decay rates.

The paper is organized as follows. Section \ref{Sect-Model} introduces the
non-Hermitian JCM, with the exact solution given in Section \ref%
{Sect-Solution}. Section \ref{Sect-spin-texture} analytically presents the
spin texture in the parity symmetry. Section \ref{Sect-invariant-nodes} shows
the invariant nodes in the spin texture. Section \ref{Sect-wind} figures out
the spin winding number and the winding direction in different planes. Section %
\ref{Sect-Tilting-Angle} gives the tilting angle of the spin winding plane
in the non-Hermiticity. Section \ref{Sect-Reversal-Gap-Closing} reveals the
reversal transition of spin winding at gap closing. Section \ref%
{Sect-Reversal-No-Gap-Closing} unveils the level-independent reversal
transition without gap closing. Section \ref{Sect-Super-Invariant} tracks
out the level-independent super-invariant point with untilted winding angle
also in a gapped situation. Reversals of spin winding directions are more straightly demonstrated for the three special points together in Section \ref{Sect-Winding-Direction-transition}. Section \ref{Sect-3D-Boundaries} gives overviews in 3D boundaries.
 Section \ref{Sect-Conclusions} makes a summary
with conclusions and discussions.

\section{Non-Hermitian Jaynes-Cummings Model}

\label{Sect-Model}

We consider a general non-Hermitian JCM~\cite%
{NonHermitianJCM-2022SciChina,JC-model,JC-Larson2021}%
\begin{equation}
H=\widetilde{\omega }a^{\dagger }a+\frac{\widetilde{\Omega }}{2}\sigma _{x}+%
\widetilde{g}\left( \widetilde{\sigma }_{-}a^{\dagger }+\widetilde{\sigma }%
_{+}a\right)  \label{H}
\end{equation}%
where the parameters are complex\cite{NonHermitianJCM-2022SciChina}%
\begin{equation}
\widetilde{\omega }=\omega -i\kappa ,\quad \widetilde{\Omega }=\Omega
-i\gamma ,\quad \widetilde{g}=g-i\Gamma .
\end{equation}%
The Hamiltonian $H$ describes the coupling between a bosonic mode with
frequency $\omega $ and a qubit with level splitting $\Omega $. The boson is
created (annihilated) by $a^{\dagger }$ ($a)$ and the qubit is represented
by the Pauli matrices $\sigma _{x,y,z}$. while the coupling strength is
denoted by $g$. The non-Hermiticity stems from $\kappa$, $\gamma $ and $%
\Gamma $, which may be effectively induced by the dissipation and decay
rates of the boson cavity, the qubit and the coupling.~\cite%
{NonHermitianJCM-2022SciChina} The steady-state features of dissipative
systems in Lindblad formalism can be well achieved by the non-Hermitian
Hamiltonian with neglected quantum jump terms~\cite%
{Plenio1998-quantum-jump,NonHermitian-Nori2019}, as also numerically
confirmed~\cite{NonHermitianJCM-2022SciChina} for the JCM considered in this
work. Indeed, in such a situation, the Liouvillian ${\cal {L}}$ in the
Lindblad master equation for dynamics and the non-Hermitian Hamiltonian $H$
have the same eigenvectors~\cite{NonHermitian-Nori2019} in the sense
that the eigenmatrices of ${\cal {L}}$ are constructed by the eigenvectors
of $H$. Setting $\kappa ,\gamma ,\Gamma =0$ one retrieves the conventional
JCM~\cite{JC-model,JC-Larson2021}. In the present work, rather than
considering the dynamics we focus on the topological features of the
eigenstates of the non-Hermitian Hamiltonian. Here in (\ref{H}) we have
adopted the spin notation as in ref.\cite{Irish2014}, in which $\sigma
_{z}=\pm $ conveniently represents the two flux states in the flux-qubit
circuit system~\cite{flux-qubit-Mooij-1999}, and $\widetilde{ \sigma }^{\pm
}=(\sigma _{z}\mp i\sigma _{y})/2$ raises and lowers the spin on $\sigma
_{x} $ basis. One can return to the conventional notation on $\sigma _{z}$
basis via a spin rotation \{$\sigma _{x},\sigma _{y},\sigma _{z} $\} $%
\rightarrow $ \{$\sigma _{z},-\sigma _{y},\sigma _{x}$\} around the axis $%
\vec{x}+\vec{z}$.

The topological feature of the JCM will manifest itself in the position
space where the Hamiltonian can be rewritten as
\begin{eqnarray}
&&H=\frac{\widetilde{\omega }}{2}\hat{p}^{2}+\widetilde{v}_{\sigma
_{z}}\left( x\right) +H_{+}\sigma _{+}+H_{-}\sigma _{-},  \label{Hx} \\
&&H_{\pm }=\frac{\widetilde{\Omega }}{2}\mp \widetilde{g}_{y}i\sqrt{2}\hat{p}%
,  \label{Hpm}
\end{eqnarray}%
by the transformation $a^{\dagger }=(\hat{x}-i\hat{p})/\sqrt{2},$ $a=(\hat{x}%
+i\hat{p})/\sqrt{2}$ with the position $x$ and the momentum $\hat{p}=-i\frac{%
\partial }{\partial x}$. Now the spin raising and lowering operators, $%
\sigma _{\pm }$ without tildes as redefined by $\sigma _{x}=\sigma
_{+}+\sigma _{-}$ and $\sigma _{y}=-i(\sigma _{+}-\sigma _{-})$, are acting
on $\sigma _{z}=\pm $ basis. In such a representation $\widetilde{v}_{\sigma
_{z}}(x)=\widetilde{\omega }\left( x+\widetilde{g}_{z}^{\prime }\sigma
_{z}\right) ^{2}/2+\varepsilon _{0}^{z}$ is an effective spin-dependent
potential with potential displacement $\widetilde{g}_{z}^{\prime }=\sqrt{2}%
\widetilde{g}_{z}/\omega $ away from the origin $x=0$, where $\widetilde{g}%
_{z}=\frac{1}{2}\widetilde{g}$ and $\varepsilon _{0}^{z}=-\frac{1}{2}[%
\widetilde{g}_{z}^{\prime 2}+1]\widetilde{\omega }$. The $\widetilde{\Omega }
$ term now plays a role of spin flipping in the $\sigma _{z}$ space or
tunneling in the position space.\cite{Ying2015,Irish2014} In (\ref{Hpm}) we
have also defined $\widetilde{g}_{y}=\frac{1}{2}\widetilde{g}$. These $%
\widetilde{g}_{y}$ terms together can be written as $\sqrt{2}\widetilde{g}%
_{y}\hat{p}\sigma _{y}$, which actually resembles \cite{Ying-Stark-top} the
Rashba spin-orbit coupling in nanowires~\cite%
{Nagasawa2013Rings,Ying2016Ellipse,Ying2017EllipseSC,Ying2020PRR,Gentile2022NatElec} or the equal-weight mixture \cite%
{Ying-gapped-top,LinRashbaBECExp2011,LinRashbaBECExp2013Review} of the
linear Dresselhaus \cite{Dresselhaus1955} and Rashba \cite{Rashba1984}
spin-orbit couplings. This may be the origin why the JCM manifests the
topological feature of spin winding as in the nanowire systems\cite%
{Nagasawa2013Rings,Ying2016Ellipse} which we shall address later on in this
work more generally in the presence of non-Hermiticity.

\section{U(1) Symmetry and Exact Solution}

\label{Sect-Solution}

Despite in the presence of non-Hermiticity, the JCM (\ref{H}) still
possesses the $U(1)$ symmetry as the conventional Hermitian JCM, denoted by
the excitation number $\widehat{n}+\left\vert \Uparrow \right\rangle
\left\langle \Uparrow \right\vert $ or $\widehat{n}+\sigma _{x}/2+1/2$ which
commutes with the Hamiltonian. With the constraint of this $U(1)$ symmetry,
the eigenstates are composed on the bases with a same excitation number:
\begin{eqnarray}
\psi _{n}^{\left( \eta \right) } &=&\left( C_{n\Uparrow }^{\left( \eta
\right) }\left\vert n-1,\Uparrow \right\rangle +C_{n\Downarrow }^{\left(
\eta \right) }\left\vert n,\Downarrow \right\rangle \right) /\sqrt{%
N_{n}^{(\eta )}},  \label{WaveF-JC-n} \\
\psi _{0} &=&\left\vert 0,\Downarrow \right\rangle ,
\end{eqnarray}%
where $\eta =\pm $ labels two branches of energy levels, $n=1,2,\cdots $
denotes the Fock state on photon number basis and $\Uparrow ,\Downarrow $
represent the two spins states of $\sigma _{x}$. The coefficients in (\ref%
{WaveF-JC-n}) are explicitly available as
\begin{eqnarray}
C_{n\Uparrow }^{\left( \eta \right) } &=&e_{-}+\eta \sqrt{e_{-}^{2}+n\
\widetilde{g}^{2}},  \label{Cx-up} \\
C_{n\Downarrow }^{\left( \eta \right) } &=&\widetilde{g}\sqrt{n},
\label{Cx-down}
\end{eqnarray}%
where $e_{+}=\left( n-\frac{1}{2}\right) \widetilde{\omega }$, $e_{-}=\frac{1%
}{2}\left( \widetilde{\Omega }-\widetilde{\omega }\right) $ and $%
N_{n}^{\left( \eta \right) }=|C_{n\Uparrow }^{\left( \eta \right)
}|^{2}+|C_{n\Downarrow }^{\left( \eta \right) }|^{2}$ is the normalization
factor. Correspondingly the eigenenergies are determined by%
\begin{eqnarray}
E^{\left( n,\eta \right) } &=&e_{+}+\eta \sqrt{e_{-}^{2}+n\ \widetilde{g}^{2}%
},  \label{E-n-JC} \\
E^{0} &=&-\frac{\widetilde{\Omega }}{2}.
\end{eqnarray}%
With the explicit exact solution on hand, we can discuss the topological
feature analytically and rigorously as in the following sections.

\section{Parity Symmetry and Spin Texture}

\label{Sect-spin-texture}


\begin{figure*}[t]
\includegraphics[width=2.0\columnwidth]{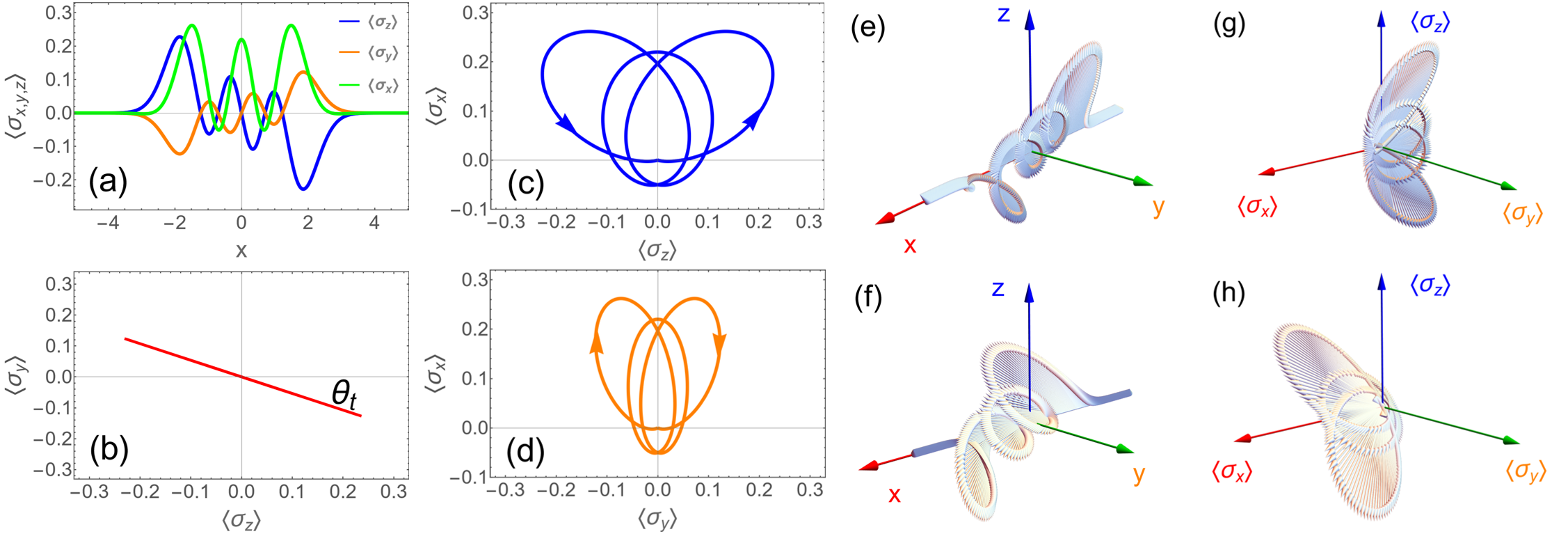}
\caption{Spin texture and spin winding with and without non-Hermiticity.
a) Space dependence of $\langle \sigma _{x} \rangle $, $\langle \sigma _{y} \rangle $ and $\langle \sigma _{z} \rangle $.
b) Spin projection in the $\langle \sigma _{z} \rangle $-$\langle \sigma _{y} \rangle $ plane and tilting angle $\theta _{\rm t}$ of the spin-winding plane.
c) Spin winding in the $\langle \sigma _{z} \rangle $-$\langle \sigma _{x} \rangle $ plane.
d) Spin winding in the $\langle \sigma _{y} \rangle $-$\langle \sigma _{x} \rangle $ plane.
e) 3D Spin texture without non-Hermiticity.
f) 3D Spin texture with non-Hermiticity.
g) 3D Spin winding without non-Hermiticity.
h) 3D Spin texture with non-Hermiticity.
Here in all panels $g=0.1g_{\rm s}$, $\omega=0.9$, $n=3$, $\eta=-1$, with $\Omega=1$ as the unit. The non-Hermiticity parameters $\kappa =0.5$, $\gamma =0.2$, $\Gamma =0.1$ in (a-d,f,h), while $\kappa =\gamma =\Gamma =0$ in (e,g).
}
\label{Fig-Spin-Winding}
\end{figure*}
\begin{figure*}[t]
\includegraphics[width=2.0\columnwidth]{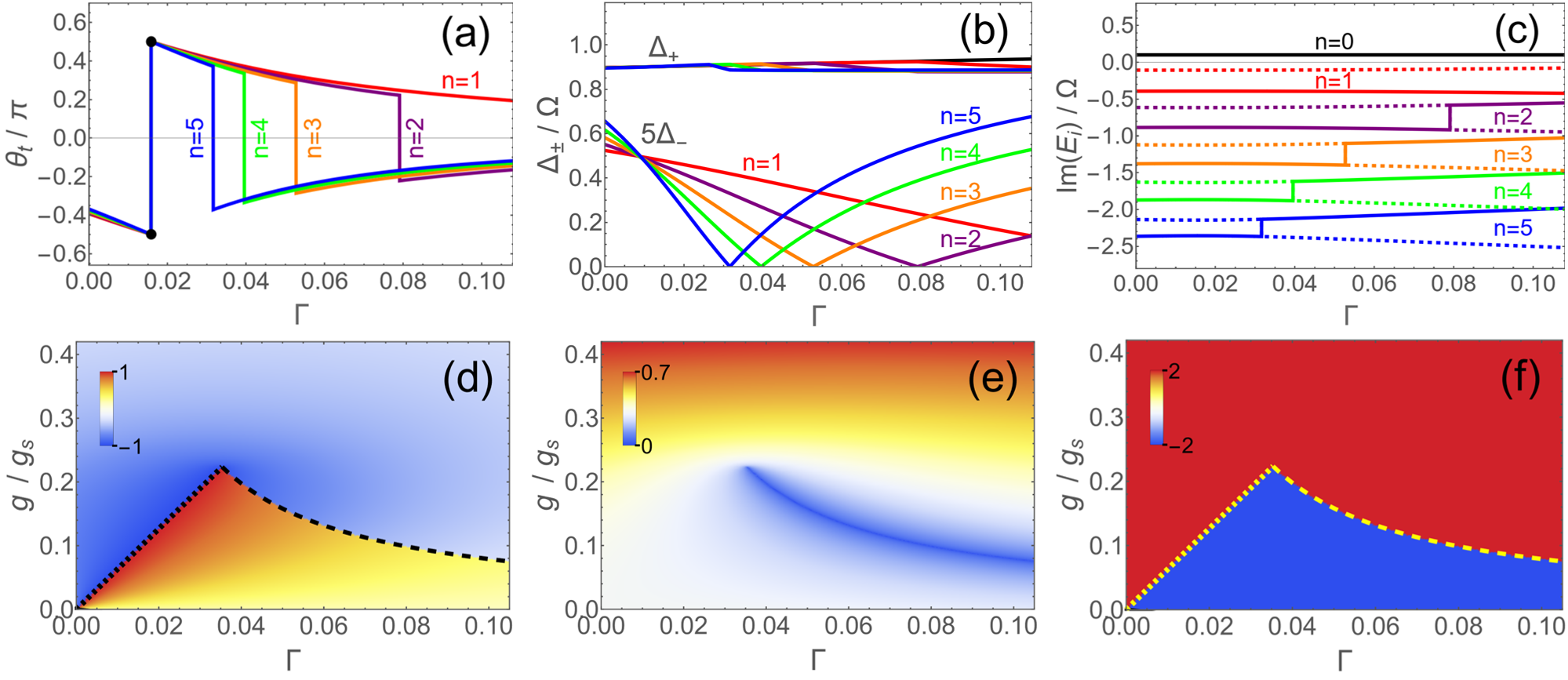}
\caption{Reversal transitions of tilting angle and spin-winding direction with gap closing and without gap closing. a-c) Tilting angle $\theta _{\rm t}$ (a), energy gap $\Delta_{\pm}$ and imaginary part of $E_i$ versus $\Gamma$ at fixed $g=0.1g_{\rm s}$. d-f) The scaled tilting angle $\theta _{\rm t}/(\pi/2)$ (d), the smaller gap $\Delta_{-}$ plotted by $\Delta_{-}^{1/2}$ (e) and the spin winding number $n_{\rm w}^{zx}$ (f) in the $\Gamma$-$g$ plane.  Here in all panels $\eta=-1$, $\kappa =0.5$, $\gamma =0.2$, $\omega=0.9$, with $\Omega=1$ as the unit.  $n=1,2,3,4,5$ in (a-c) and  $n=2$ in (d-f). The dotted lines and the dashed lines in (d,f) are analytic $g_{\rm R}$ and $g_{\rm GR}$ in (\ref{gR-1}) and (\ref{gSR-1}).
}
\label{Fig-Reverse-Transition}
\end{figure*}

Corresponding to the Hamitonian (\ref{Hx}), the eigenstate in (\ref%
{WaveF-JC-n}) can be represented as eigen wave function in the position
space
\begin{eqnarray}
\psi _{+}^{x}\left( x\right) &\equiv &\psi _{n,\Uparrow }^{\left( x,\eta
\right) }\left( x\right) =C_{n\Uparrow }^{\left( \eta \right) }\phi
_{n-1}\left( x\right) /\sqrt{N_{n}^{(\eta )}},  \label{wave-X-1} \\
\psi _{-}^{x}\left( x\right) &\equiv &\psi _{n,\Downarrow }^{\left( x,\eta
\right) }\left( x\right) =C_{n\Downarrow }^{\left( \eta \right) }\phi
_{n}\left( x\right) /\sqrt{N_{n}^{(\eta )}}.  \label{wave-X-2}
\end{eqnarray}%
where $\phi _{n}\left( x\right) =\langle x|n\rangle =\frac{1}{\pi ^{1/4}%
\sqrt{2^{n}n!}}H_{n}\left( x\right) e^{-x^{2}/2}$ with the Hermite
polynomial $H_{n}\left( x\right) $ is the eigen function of the quantum
harmonic oscillator with quantum number $n.$ As afore-mentioned the spin
notation in the Hamiltonian (\ref{H}) and the eigen state (\ref{WaveF-JC-n})
is on the spin-$\sigma _{x}$ basis denoted by $\Uparrow $ and $\Downarrow $,
we can also transform onto the spin-$\sigma _{z}$ basis represented by $%
\uparrow $ and $\downarrow $, via the basis transformation%
\begin{eqnarray}
\left\vert \uparrow \right\rangle &=&\frac{\left\vert \Uparrow \right\rangle
+\left\vert \Downarrow \right\rangle }{\sqrt{2}},\qquad \left\vert
\downarrow \right\rangle =\frac{\left\vert \Uparrow \right\rangle
-\left\vert \Downarrow \right\rangle }{\sqrt{2}}, \\
\left\vert \Uparrow \right\rangle &=&\frac{\left\vert \uparrow \right\rangle
+\left\vert \downarrow \right\rangle }{\sqrt{2}},\qquad \left\vert
\Downarrow \right\rangle =\frac{\left\vert \uparrow \right\rangle
-\left\vert \downarrow \right\rangle }{\sqrt{2}}.
\end{eqnarray}%
On the $\sigma _{z}$ basis the eigen wave function becomes
\begin{equation}
\psi _{n}^{\left( z,\eta \right) }=\ \psi _{+}^{z}\left( x\right) \left\vert
\uparrow \right\rangle +\psi _{-}^{z}\left( x\right) \left\vert \downarrow
\right\rangle  \label{wave-Z-0}
\end{equation}%
with the spin components%
\begin{eqnarray}
\psi _{+}^{z}\left( x\right) &\equiv &\psi _{n,\uparrow }^{\left( z,\eta
\right) }\left( x\right) =\frac{C_{n\Uparrow }^{\left( \eta \right) }\phi
_{n-1}\left( x\right) +C_{n\Downarrow }^{\left( \eta \right) }\phi
_{n}\left( x\right) }{\sqrt{2N_{n}^{(\eta )}}},  \label{wave-Z-1} \\
\psi _{-}^{z}\left( x\right) &\equiv &\psi _{n,\downarrow }^{\left( z,\eta
\right) }\left( x\right) =\frac{C_{n\Uparrow }^{\left( \eta \right) }\phi
_{n-1}\left( x\right) -C_{n\Downarrow }^{\left( \eta \right) }\phi
_{n}\left( x\right) }{\sqrt{2N_{n}^{(\eta )}}}.  \label{wave-Z-2}
\end{eqnarray}
From the above spin components of the eigen wave function on the $\sigma
_{x} $ and $\sigma _{z}$ bases, we can determine the spin texture by
\begin{eqnarray}
\langle \sigma _{z}\left( x\right) \rangle &=&\left\vert \psi _{+}^{z}\left(
x\right) \right\vert ^{2}-\left\vert \psi _{-}^{z}\left( x\right)
\right\vert ^{2}  \label{SpinZ-byWave} \\
\langle \sigma _{x}\left( x\right) \rangle &=&\left\vert \psi _{+}^{x}\left(
x\right) \right\vert ^{2}-\left\vert \psi _{-}^{x}\left( x\right)
\right\vert ^{2},  \label{SpinX-byWave} \\
\langle \sigma _{y}\left( x\right) \rangle &=&i\left[ \psi _{-}^{z}\left(
x\right) ^{\ast }\psi _{+}^{z}\left( x\right) -\psi _{+}^{z}\left( x\right)
^{\ast }\psi _{-}^{z}\left( x\right) \right] .  \label{SpinY-byWave}
\end{eqnarray}

Note that the non-Hermitian model $H$ also possesses the parity symmetry as
the Hermitian case, i.e. the parity operator $\hat{P}=\sigma _{x}\left(
-1\right) ^{a^{\dagger }a}$ commutes with the Hamiltonian, $[\hat{P},H]=0$
even in the presence of non-Hermiticity. An eigenstate has a negative parity
when $n$ is even, while the parity is positive for odd $n$:
\begin{equation}
\hat{P}\psi _{n}^{\left( x,\eta \right) }=\left( -1\right) ^{n-1}\psi
_{n}^{\left( x,\eta \right) },\quad \hat{P}\psi _{0}=\left( -1\right) \psi
_{0},  \label{parity-wave-x}
\end{equation}%
as $\left( -1\right) ^{a^{\dagger }a}$ of $\hat{P}$ in the position
representation inverses the space: $x\rightarrow -x$.\cite%
{Ying2020-nonlinear-bias} On the $\sigma _{z}$ basis the parity involves
both spin reversal and space inversion
\begin{equation}
\psi _{n,\uparrow }^{\left( z,\eta \right) }\left( x\right) =\left(
-1\right) ^{n-1}\psi _{n,\downarrow }^{\left( z,\eta \right) }\left(
-x\right) .  \label{wave-Z-Inverse-x}
\end{equation}%
The parity symmetry of the eigen wave function leads to symmetric and
anti-symmetric properties of spin texture:%
\begin{eqnarray}
\langle \sigma _{x}\left( -x\right) \rangle &=&\langle \sigma _{x}\left(
x\right) \rangle , \\
\langle \sigma _{z}\left( -x\right) \rangle &=&-\langle \sigma _{z}\left(
x\right) \rangle , \\
\langle \sigma _{y}\left( -x\right) \rangle &=&-\langle \sigma _{y}\left(
x\right) \rangle ,
\end{eqnarray}%
which actually guarantee the close form of spin winding to form the
topological feature, as will be discussed in Section \ref{Sect-wind}.

We can decompose the wave-function coefficients into real and imaginary parts
\begin{eqnarray}
C_{n\Uparrow }^{\left( \eta ,
\mathop{\rm Re}
\right) } &=&\frac{1}{2}\left( \Omega -\omega \right) +\eta R\cos \frac{
\vartheta }{2}, \\
C_{n\Uparrow }^{\left( \eta ,
\mathop{\rm Im}
\right) } &=&\frac{1}{2}\left( \kappa -\gamma \right) +\eta R\sin \frac{
\vartheta }{2}, \\
C_{n\Downarrow }^{\left( \eta ,
\mathop{\rm Re}
\right) } &=&g\sqrt{n},\qquad C_{n\Downarrow }^{\left( \eta ,
\mathop{\rm Im}
\right) }=-\Gamma \sqrt{n},
\end{eqnarray}%
where%
\begin{eqnarray}
R &=&\left( A^{2}+B^{2}\right) ^{1/4},\quad \vartheta =\arg \left(
A-iB\right) ,  \label{Def-theta} \\
A &=&n\ \left( g^{2}-\Gamma ^{2}\right) +\frac{1}{4}d_{\Omega \omega }^{2}-
\frac{1}{4}d_{\kappa \gamma }^{2},  \label{Def-A} \\
B &=&2n\ g\ \Gamma -\frac{1}{2}d_{\kappa \gamma }d_{\Omega \omega },
\label{Def-B}
\end{eqnarray}%
with $d_{\kappa \gamma }=\kappa -\gamma $, $d_{\Omega \omega }=\Omega
-\omega $. Thus the spin texture for state $\psi _{n}^{\left( x,\eta \right)
}$ can be analytically obtained to be%
\begin{eqnarray}
&&\langle \sigma _{z}\left( x\right) \rangle =\frac{e^{-x^{2}}\ \widetilde{C}
_{z}}{2^{n-3/2}\widetilde{N}_{\sigma }}H_{n-1}\left( x\right) H_{n}\left(
x\right) ,  \label{SpinZ-expression} \\
&&\langle \sigma _{x}\left( x\right) \rangle =\frac{e^{-x^{2}}}{2^{n}
\widetilde{N}_{\sigma }}[\widetilde{D}_{x}H_{n-1}\left( x\right)
^{2}-2\left( g^{2}+\Gamma ^{2}\right) H_{n}\left( x\right) ^{2}],
\label{SpinX-expression} \\
&&\langle \sigma _{y}\left( x\right) \rangle =\frac{e^{-x^{2}}\ \widetilde{C}
_{y}}{2^{n-3/2}\widetilde{N}_{\sigma }}H_{n-1}\left( x\right) H_{n}\left(
x\right) ,  \label{SpinY-expression}
\end{eqnarray}
where the coefficients before the Hermite polynomials are defined as
\begin{eqnarray}
\widetilde{C}_{z} &=&gd_{\Omega \omega }-\Gamma d_{\kappa \gamma }+2\eta
R(g\cos \frac{\vartheta }{2}-\Gamma \sin \frac{\vartheta }{2}),
\label{Coefficient-z} \\
\widetilde{C}_{y} &=&\Gamma d_{\Omega \omega }+gd_{\kappa \gamma }+2\eta
R(\Gamma \cos \frac{\vartheta }{2}+g\sin \frac{\vartheta }{2}),
\label{Coefficient-y} \\
\widetilde{D}_{x} &=&d_{\Omega \omega }^{2}+d_{\kappa \gamma
}^{2}+4R^{2}+4\eta R[d_{\Omega \omega }\cos \frac{\vartheta }{2}+d_{\kappa
\gamma }\sin \frac{\vartheta }{2}],  \label{Coefficient-x}
\end{eqnarray}%
and $\widetilde{N}_{\sigma }=\sqrt{\pi }(n-1)!\ 2N_{n}^{\left( \eta \right)
} $. In the conventional Hermitian JCM, we have $\kappa =\gamma =\Gamma
=0,B=0$ and $\vartheta =0$, which lead to $\langle \sigma _{x}\left(
x\right) \rangle =0$.\cite{Ying-JC-winding} In contrast, we now have a
finite $\langle \sigma _{y}\left( x\right) \rangle $ contribution in the
presence of the non-Hermiticity, which induces an out-of-plane component
apart from the spin winding in the $\langle \sigma _{z}\left( x\right)
\rangle $-$\langle \sigma _{x}\left( x\right) \rangle $ plane as will be
discussed in more detail in Section \ref{Sect-wind}. For state $\psi _{0}$,
we have $\langle \sigma _{z}\left( x\right) \rangle =\langle \sigma
_{y}\left( x\right) \rangle =0$ and $\langle \sigma _{x}\left( x\right)
\rangle =-e^{-x^{2}}/\sqrt{\pi }$.

\section{Invariant Nodes in Spin Texture}

\label{Sect-invariant-nodes}

An illustration of the spin texture is given in {\bf Figure} \ref%
{Fig-Spin-Winding}a for $\langle \sigma _{x}\left( x\right) \rangle $
(green), $\langle \sigma _{y}\left( x\right) \rangle $ (orange) and $\langle
\sigma _{z}\left( x\right) \rangle $ (blue). One sees that the spin texture
has some oscillations which give rises to some nodes $x_{Z,i}^{\alpha }$
where $\langle \sigma _{\alpha }\left( x_{Z,i}^{\alpha }\right) \rangle =0$
for $\alpha \in \{x,y,z\}$. It has been noticed that the situation of nodes
forms a topological feature that characterizes the topological phases in
light-matter interaction.\cite%
{Ying-2021-AQT,Ying-gapped-top,Ying-Stark-top,Ying-Spin-Winding,Ying-JC-winding}
With a fixed node number one cannot go to another node state by continuous
shape deformation, just as one cannot change a torus into a sphere by a
continuous deformation in the topological picture of so-called rubber-sheet
geometry. It has also been found that nodes have a correspondence to spin
winding \cite{Ying-Spin-Winding,Ying-JC-winding} which is a more physical
topological feature. The node sorting can algebraically decode the
topological information encoded geometrically in the wave functions and spin
windings.\cite{Ying-Spin-Winding} Before addressing the spin winding in next
section, here it is worthwhile to mention the invariant nodes in the spin
texture. Indeed, Equations (\ref{SpinZ-expression}) and (\ref%
{SpinY-expression}) indicate that the nodes of $\langle \sigma _{z}\left(
x\right) \rangle $ and $\langle \sigma _{y}\left( x\right) \rangle $ are the
roots of the Hermite polynomials which are independent of the system
parameters. There are $2n$\ nodes in $\langle \sigma _{x}\left( x\right)
\rangle $ and $2n-1$ nodes in $\langle \sigma _{z}\left( x\right) \rangle $
and $\langle \sigma _{y}\left( x\right) \rangle $ according to the Hermite
polynomials they contain. Involving the same Hermite polynomial factors $%
H_{n-1}\left( x\right) \ $and $H_{n}\left( x\right) $, $\langle \sigma
_{z}\left( x\right) \rangle $ and $\langle \sigma _{y}\left( x\right)
\rangle $ share the same nodes. Moreover, they are invariant not only in the
numbers but also in the positions, which may much reduce the experimental
cost and simplify the identification of the topological states in
experimental simulations, while usually in condensed matter one needs
measurements over a global space to exactly identify the topological state
\cite%
{MajoranaExpScience-2014,MajoranaExpScience-2017,ChernNumberExpPRL-2021,TopCriterion}
as topological feature is a global property rather than a local order
parameter in the traditional phase transitions. Here, we see that the
invariant nodes are unaffected by the non-Hermiticity.

\section{Spin Winding}

\label{Sect-wind}

The spin texture is unfolded in the position space, while it manifests
itself to be spin winding in spin expectation planes. In Figure \ref%
{Fig-Spin-Winding}b,c, the evolution of local spin expectation forms winding
around the origin with a close form in the $\langle \sigma _{z}\left(
x\right) \rangle $-$\langle \sigma _{x}\left( x\right) \rangle $ plane
(panel (c)) and the $\langle \sigma _{y}\left( x\right) \rangle $-$\langle
\sigma _{x}\left( x\right) \rangle $ plane (panel (d)). Such a close form of
spin winding is guaranteed by the parity symmetry, as mentioned in Section %
\ref{Sect-spin-texture}. The two key factors to characterize the spin
winding are the spin winding number and the winding direction, which we
shall figure out explicitly in this section.

\begin{figure*}[t]
\includegraphics[width=2.0\columnwidth]{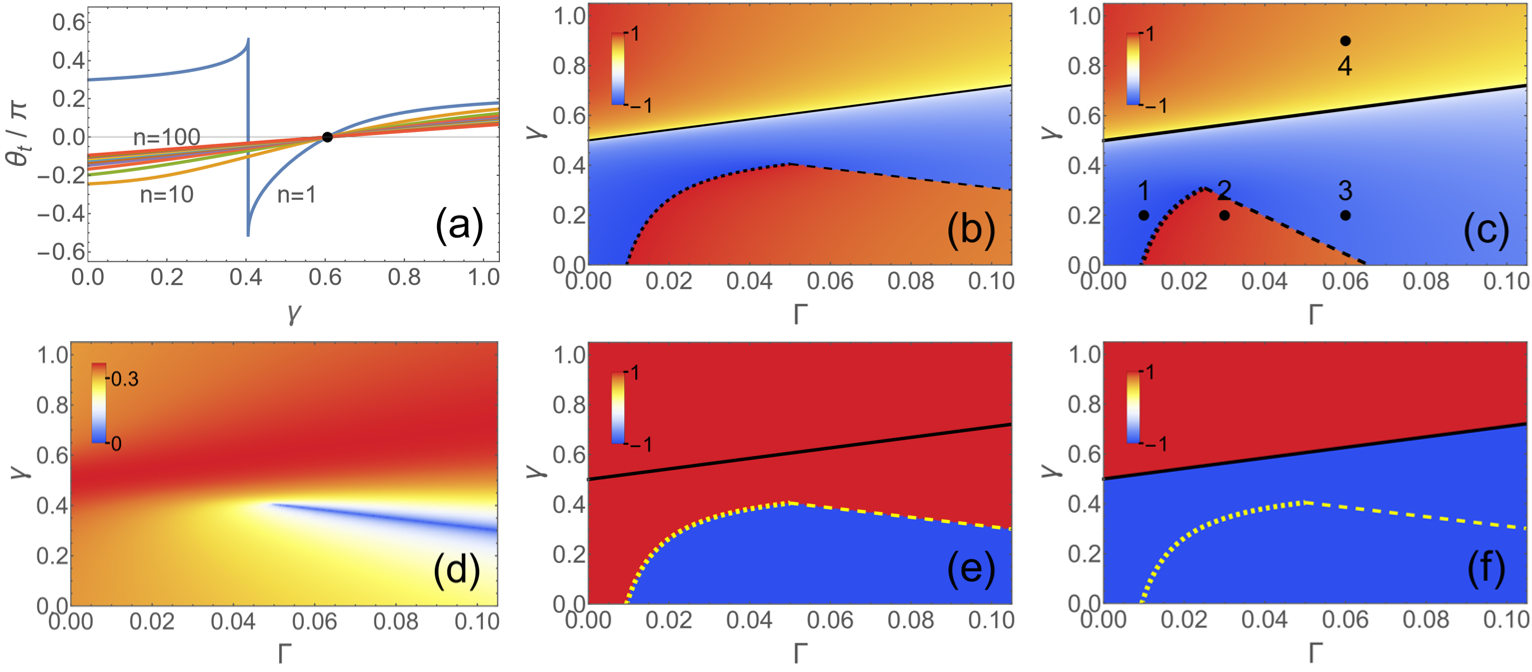}
\caption{Super-invariant point for un-induced spin tilting by non-Hermiticity and for all energy levels. a) Spin tilting angle $\theta _{\rm t}$ versus $\gamma$ for different levels (every 10 levels plotted up to $n=100$) at fixed $\Gamma =0.05$. The dot marks the super-invariant point. b-f) $\theta _{\rm t}$ plotted by $(2\theta _{\rm t}/\pi)^{1/4}$ (b,c), $\Delta_{-}^{1/2}$ (d), $n^{zx}_{\rm w}$ (e) and $n^{yx}_{\rm w}$ (f) in the $\Gamma$-$\gamma$ plane. Here in all panels $\eta=-1$, $\kappa =0.5$, $g =0.1g_{\rm s}$ and $\omega=0.9$, with $\Omega=1$ as the unit. $\gamma =0.05$ in (a), $n=1$ in (b,d-f) and $n=4$ in (c). The the dashed lines, dotted lines and solid lines in (b,c,e,f) are analytic $\gamma_{\rm R}$ and $\gamma_{\rm GR}$ in (\ref{gR-2}) (\ref{gSR-2}) and (\ref{gSI-2}). The dots in (c) mark some representative points for the plots of spin winding direction in Figure \ref{Fig-Winding-direction}.
}
\label{Fig-Super-Invariant}
\end{figure*}
\begin{figure*}[t]
\includegraphics[width=2\columnwidth]{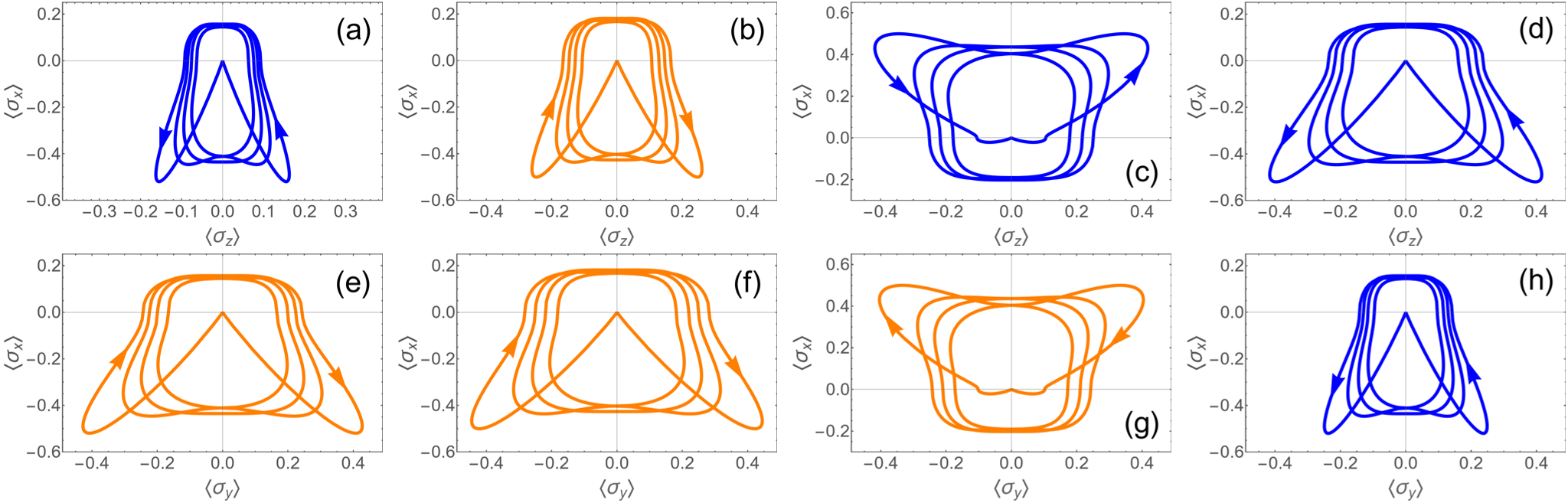}
\caption{Winding direction reversals: Spin winding in
the $\langle \sigma _{z}\left( x\right) \rangle $-$\langle \sigma _{x}\left( x\right) \rangle $ plane (a-d)
and
the $\langle \sigma _{y}\left( x\right) \rangle $-$\langle \sigma _{x}\left( x\right) \rangle $ plane (e-h), respectively
before the reversal transition (a,e), after the reversal transition and before the gapped reversal transition (b,f), after the gapped reversal transition (c,g) and beyond the super-invariant point (d,h), corresponding to the dots marked by $1,2,3,4$ as representative points of the phases in Figure \ref{Fig-Super-Invariant}c.  a,e) $\Gamma =0.01$ and $\gamma =0.2$;  b,f) $\Gamma =0.03$ and $\gamma =0.2$;  c,g) $\Gamma =0.06$ and $\gamma =0.2$;  d,h) $\Gamma =0.06$ and $\gamma =0.9$.  Here in all panels $n=4$, $\eta=-1$, $\kappa =0.5$, $\gamma =0.2$, $\omega=0.9$, $g=0.1g_{\rm s}$ with $\Omega=1$ as the unit. The spin amplitudes are plotted by $|\langle \sigma _{\alpha}\left( x\right) \rangle |^{1/2}$ for all components $\alpha=x,y,z$ to amplify the detail. The spin winding direction is indicated by the arrows, and also represented by the color with blue for counter-clockwise and orange for clockwise.
}
\label{Fig-Winding-direction}
\end{figure*}

\subsection{Spin Winding Number}

\label{Sect-wind-No}

The spin winding number in the $\langle \sigma _{\alpha }\left( x\right)
\rangle $-$\langle \sigma _{\beta }\left( x\right) \rangle $ plane, where $%
\alpha ,\beta \in \{x,y,z\},$ is defined by%
\begin{equation}
n_{{\rm w}}^{\alpha \beta }=\frac{1}{2\pi }\int_{-\infty }^{\infty }\frac{%
\langle \sigma _{\alpha }\left( x\right) \rangle \partial _{x}\langle \sigma
_{\beta }\left( x\right) \rangle -\langle \sigma _{\beta }\left( x\right)
\rangle \partial _{x}\langle \sigma _{\alpha }\left( x\right) \rangle }{%
\langle \sigma _{\alpha }\left( x\right) \rangle ^{2}+\langle \sigma _{\beta
}\left( x\right) \rangle ^{2}}dx,  \label{nW-by-integration}
\end{equation}%
which has also been applied in topological classification in nanowire
systems and quantum systems with geometric driving.\cite%
{Ying2016Ellipse,Ying2017EllipseSC,Ying2020PRR,Gentile2022NatElec} Although
Equation (\ref{nW-by-integration}) involves calculus of both integral and
differential which is numerically more difficult to treat, it was pointed
out that the winding number can be extracted algebraically in terms of the
nodes without integral or differential.\cite%
{Ying-Spin-Winding,Ying-JC-winding} Indeed, by assuming $M_{\alpha }$ number
of $\langle \sigma _{\alpha }\left( x\right) \rangle $ nodes at finite node
position $x_{Z,i}^{\alpha }$, it has been rigorously proven\cite%
{Ying-JC-winding} that the spin winding number defined by the involved
geometric integral (\ref{nW-by-integration}) is equal to the simple
algebraic sum of spin signs at nodes
\begin{eqnarray}
n_{{\rm w}}^{\alpha \beta } &=&-\sum_{i=0}^{M_{\beta }}\frac{sgn\langle
\sigma _{\alpha }(x_{Z,i+1}^{\beta })\rangle -sgn\langle \sigma _{\alpha
}(x_{Z,i}^{\beta })\rangle }{4\eta _{\beta }\left( i\right) }
\label{nW-sigma-alpha} \\
&=&\sum_{i=0}^{M_{\alpha }}\frac{sgn\langle \sigma _{\beta
}(x_{Z,i+1}^{\alpha })\rangle -sgn\langle \sigma _{\beta }(x_{Z,i}^{\alpha
})\rangle }{4\eta _{\alpha }\left( i\right) }  \label{nW-sigma-beta}
\end{eqnarray}%
which holds for a generic spin winding. Here $\eta _{\alpha }\left( i\right)
$ is the sign of $\langle \sigma _{\alpha }\left( x\right) \rangle $ in
space section $x\in (x_{Z,i}^{\alpha },x_{Z,i+1}^{\alpha }).$ The edge
sections $i=0,M_{\alpha }$ are $(-\infty ,x_{Z,1}^{\alpha })$ and $%
(x_{Z,M_{\alpha }}^{\alpha },\infty )$ by setting $x_{Z,0}^{\alpha }=-\infty
$ and $x_{Z,M_{\alpha }+1}^{\alpha }=\infty $. For the nodes $sgn\langle
\sigma _{\alpha }\rangle ={\rm sign}\langle \sigma _{\alpha }\rangle ,$
while for the infinity ends $sgn\langle \sigma _{\alpha }\rangle =2\arcsin
\langle \overline{\sigma }_{\alpha }\rangle /\pi $ where $\langle \overline{%
\sigma }_{\alpha }\rangle =\langle \sigma _{\alpha }\rangle /\sqrt{\langle
\sigma _{\alpha }\rangle ^{2}+\langle \sigma _{\beta }\rangle ^{2}}$.
Despite $M_{x}\neq M_{z},$ Equations (\ref{nW-sigma-alpha}) and (\ref%
{nW-sigma-beta}) are equal as $M_{x}=2n$ with $sgn\langle \sigma _{z,y}(\pm
\infty )\rangle =0$ while $M_{z,y}=2n-1$ but with $\left\vert sgn\langle
\sigma _{x}(\pm \infty )\rangle \right\vert =1$. The difference of $%
sgn\langle \sigma _{z,y}(\pm \infty )\rangle $ and $\left\vert sgn\langle
\sigma _{x}(\pm \infty )\rangle \right\vert $ comes from the ratio limit $%
\frac{\langle \sigma _{z,y}\left( x\right) \rangle }{\langle \sigma
_{x}\left( x\right) \rangle }|_{x\rightarrow \pm \infty }\rightarrow 0 $ as $%
\langle \sigma _{x}\left( x\right) \rangle $ contains a $H_{n}\left(
x\right) ^{2}$ term which is in a higher rank of polynomial than $%
H_{n-1}\left( x\right) H_{n}\left( x\right) $ in $\langle \sigma
_{z,y}\left( x\right) \rangle .$

As all the $\langle \sigma _{\alpha }\rangle $ components in (\ref%
{SpinZ-expression})-(\ref{SpinY-expression}) only involve neighboring
Hermite polynomials, the spin is winding in one direction without
anti-winding nodes or returning knots,\cite{Ying-JC-winding} $\eta _{\alpha
}\left( i\right) $ changes the sign alternatively and brings all the nodes
into full contributions. Thus, we have the absolute spin winding numbers in
the $\langle \sigma _{z}\left( x\right) \rangle $-$\langle \sigma _{x}\left(
x\right) \rangle $ plane and the $\langle \sigma _{y}\left( x\right) \rangle
$-$\langle \sigma _{x}\left( x\right) \rangle $ plane both equal to the
excitation number%
\begin{equation}
\left\vert n_{{\rm w}}^{zx}\right\vert =\left\vert n_{{\rm w}%
}^{yx}\right\vert =n
\end{equation}%
for the eigen state. Thus the excitation number $n$ is now endowed the
topological connotation to be the topological quantum number of spin winding
not only in the conventional Hermitian JCM but also in the general
non-Hermitian JCM.

\subsection{Spin Winding Direction}

\label{Sect-wind-direction}

Besides the winding-number magnitude, the spin winding direction is also an
important quantum character. Since the spin winding has no returning within
an eigen state,\cite{Ying-JC-winding} the winding direction can be
conveniently figured out from the ratio sign of $\langle \sigma _{z,y}\left(
x\right) \rangle $ and $\langle \sigma _{x}\left( x\right) \rangle $ at
infinity where the winding starts. Actually the winding will be clockwise
when it starts in the 1st or 3rd quadrant of the $\langle \sigma
_{z,y}\left( x\right) \rangle $-$\langle \sigma _{x}\left( x\right) \rangle $
plane, as a counter-clockwise winding would spuriously generate a $2n+1$
umber of $\langle \sigma _{z,y}\left( x\right) \rangle $ nodes exceeding the
correct node number; Otherwise in the 1st or 3rd quadrant, the winding is
counter-clockwise. Thus, the spin winding is counter-clockwise (clockwise)
if the sign of
\begin{eqnarray}
s_{{\rm w}}^{zx} &=&\rm{sign}\frac{\langle \sigma _{z}\left( x\right)
\rangle }{\langle \sigma _{x}\left( x\right) \rangle }|_{x\rightarrow
-\infty }=\rm{sign}(\widetilde{C}_{z}), \\
s_{{\rm w}}^{yx} &=&\rm{sign}\frac{\langle \sigma _{y}\left( x\right)
\rangle }{\langle \sigma _{x}\left( x\right) \rangle }|_{x\rightarrow
-\infty }=\rm{sign}(\widetilde{C}_{y}),
\end{eqnarray}%
is negative (positive). Here we have taken into account the property $%
H_{n-1}\left( x\right) H_{n}\left( x\right) <0$ at $x\rightarrow -\infty $
due to $H_{n}\left( x\right) \propto x^{n}$ there in the leading order.
Finally, the complete information of the spin winding number includes the
magnitude $n$ counting the winding rounds and the sign representing the
winding direction:
\begin{eqnarray}
n_{{\rm w}}^{zx} &=&-s_{{\rm w}}^{zx}n, \\
n_{{\rm w}}^{yx} &=&-s_{{\rm w}}^{yx}n,
\end{eqnarray}%
which defines positive $n_{{\rm w}}^{\alpha \beta }$ for counter-clockwise
winding and negative for clockwise according to (\ref{nW-by-integration}).

In the conventional Hermitian JCM,
\begin{equation}
\widetilde{C}_{z}\rightarrow g\left( \Omega -\omega \right) +\eta g\sqrt{%
\left( \Omega -\omega \right) ^{2}+4g^{2}n},
\end{equation}
which indicates that all states with $\eta =-1$ have a counter-clockwise
spin winding direction, while the winding direction of the states with $\eta
=+1$ is opposite. However, such a unified picture of the winding direction
is broken in the presence of non-Hermiticity, as we shall address in
Sections \ref{Sect-Reversal-Gap-Closing}-\ref{Sect-Super-Invariant}.

\section{Tilting Angle of Spin Winding Plane}

\label{Sect-Tilting-Angle}

We have seen below (\ref{Coefficient-z}) that there is no $\langle \sigma
_{y}\left( x\right) \rangle $ component in the conventional Hermitian JCM. A
finite $\langle \sigma _{y}\left( x\right) \rangle $ component emerges in
the presence of the non-Hermiticity. Equations (\ref{SpinZ-expression})-(\ref%
{SpinY-expression}) shows that $\langle \sigma _{y}\left( x\right) \rangle $
is proportional to $\langle \sigma _{z}\left( x\right) \rangle $ and has the
same the magnitude of spin winding number, which means that the
non-Hermiticity only tilts the spin winding plane while the topological
feature is maintained. This is more clearly demonstrated by Figure \ref%
{Fig-Spin-Winding}b where the spin winding in the $\langle \sigma _{z}\left(
x\right) \rangle $-$\langle \sigma _{y}\left( x\right) \rangle $ plane is
completely flat. The zero tilting of the spin winding plane in the Hermitian
case and finite tilting in the non-Hermitian case can be further visualized
in Figure \ref{Fig-Spin-Winding}e-h by the 3D plots of the spin texture
(e,f) and spin winding (g,h), with the comparison of the Hermitian case
(e,g) and the non-Hermitian case (f,h). The tilting degree can be described
by the ratio
\begin{equation}
R_{yz}=\frac{\langle \sigma _{y}\left( x\right) \rangle }{\langle \sigma
_{z}\left( x\right) \rangle }=\frac{\widetilde{C}_{y}}{\widetilde{C}_{z}}%
=\tan \theta _{{\rm t}}
\end{equation}%
which is exactly the same at any position $x$. The tilting angle is then
denoted by%
\begin{equation}
\theta _{{\rm t}}=\arctan \frac{\widetilde{C}_{y}}{\widetilde{C}_{z}}. \label{def-theta-t}
\end{equation}

The tilting angle $\theta _{{\rm t}}$ not only describes the tilting degree
but also can help to track and reveal some winding reversal transitions and
super-invariant points, as unveiled in the following sections.


\section{Fractional Phase Gain and Reversal Transition of Spin Winding at Gap Closing}
\label{Sect-Reversal-Gap-Closing}

We find a reversal transition of the spin winding induced by the
non-Hermiticity, with the reversal both in the tilting angle of the spin winding plane and the direction of the spin winding. An example is illustrated in {\bf Figure} \ref%
{Fig-Reverse-Transition}a, where one finds a jump or a reversal of the
tilting angle $\theta _{{\rm t}}$ in varying a non-Hermiticity parameter
such as $\Gamma $. Such a reversal transition occurs for all excited levels
with level-dependent transition point. The $\theta _{{\rm t}}$ jump
originates from the square root term in (\ref{Cx-up}), $S_{r}=\sqrt{%
e_{-}^{2}+n\ \widetilde{g}^{2}}=\sqrt{A-iB}$ with $B$ in (\ref{Def-B})
induced by the non-Hermiticity. Under a negative $A$, when $B$ is crossing
its zero point a shift from $\pi $ to $-\pi $ will take place for $\vartheta
$ in (\ref{Def-theta}). This $2\pi $ phase gain is trivial in usual case, but
here inside the square root it will fractionally contribute a $\pi $ phase 
with a sign change of $S_{r}$,
which happens to reverse the tilting angle of the spin winding plane (See the proof in Appendix \ref{Appendix-Proof-Reversal}).

To see the situation of the reversal transition, we can decompose $E^{\left(
n,\eta \right) }$ into real part and imaginary part%
\begin{eqnarray}
E_{\mathop{\rm Re}}^{\left( n,\eta \right) } &=&(n-\frac{1}{2})\omega +\eta R\cos \frac{
\vartheta }{2},  \label{ReE-by-Angle} \\
E_{\mathop{\rm Im}}^{\left( n,\eta \right) } &=&-(n-\frac{1}{2})\kappa +\eta R\sin \frac{
\vartheta }{2}.  \label{ImE-by-Angle}
\end{eqnarray}%
The real part should represent the energy difference of the levels, while
the imaginary part would decide the relaxation time to the steady state in
dynamics.\cite{Plenio1998-quantum-jump,NonHermitian-Nori2019} The $\theta _{{\rm t}}$ reversal transition
occurs at gap closing of $E_{\mathop{\rm Re}}^{\left( n,\eta \right) }$, as shown by the smaller gap
$\Delta _{-}=|E_{\mathop{\rm Re}}^{\left( n,+1\right) }-E_{\mathop{\rm Re}}^{\left( n,-1\right) }|$ in Figure \ref{Fig-Reverse-Transition}b despite
that the larger gap $\Delta _{+}$ between different-$n$ levels is finite.
The gap closing stems from the $\eta R\cos \frac{\vartheta }{2}$ term of $E_{\mathop{\rm Re}}^{\left( n,\eta \right) }$, as $\cos \frac{\vartheta }{2}=0$ for $\vartheta
=\pm \pi $ at the transition point. Across the transition there is indeed a
phase shift as indicated by the imaginary part of $E^{\left( n,\eta \right) }
$ in Figure \ref{Fig-Reverse-Transition}c.

The transition boundary can be analytically found from $B=0$ to be
\begin{eqnarray}
\kappa _{{\rm R}} &=&\gamma +\frac{4ng\Gamma }{\left( \Omega -\omega \right)
},\qquad \Gamma _{{\rm R}}=\frac{\left( \kappa -\gamma \right) \left( \Omega
-\omega \right) }{4ng},  \label{gR-1} \\
\gamma _{{\rm R}} &=&\kappa -\frac{4ng\Gamma }{\left( \Omega -\omega \right)
},\qquad g_{{\rm R}}=\frac{\left( \kappa -\gamma \right) \left( \Omega
-\omega \right) }{4n\Gamma }.  \label{gR-2}
\end{eqnarray}%
Note, as afore-mentioned, that the transition occurs only under the
condition of negative $A,$ which requires
\begin{equation}
n\ \left( g^{2}-\Gamma ^{2}\right) +\frac{1}{4}d_{\Omega \omega }^{2}-\frac{1%
}{4}d_{\kappa \gamma }^{2}<0.  \label{R-condition-gmma}
\end{equation}%
Here the condition (\ref{R-condition-gmma}) is imposed on the transition
boundary. For example (\ref{R-condition-gmma}) becomes
\begin{equation}
\Gamma _{{\rm R}}^{2}>\Gamma _{{\rm A}}^{2}=g^{2}+\frac{1}{4n}\left[ \left(
\Omega -\omega \right) ^{2}-\left( \kappa -\gamma \right) ^{2}\right]
\end{equation}%
which means that one always has the transition in varying $\Gamma $ if the
other parameters lead to $\Gamma _{{\rm A}}^{2}<0$, otherwise for $\Gamma _{%
{\rm A}}^{2}>0$ the existence of the reversal transition requires
\begin{equation}
\Gamma >\Gamma _{{\rm R}}^{\min }\equiv \frac{\left\vert \Omega -\omega
\right\vert }{2\sqrt{n}}\quad \text{and}\quad \left( \kappa -\gamma \right)
\left( \Omega -\omega \right) >0
\end{equation}%
where $\Gamma $ is supposed to be positive.

In Figure \ref{Fig-Reverse-Transition}d-f we provide the maps in the $\Gamma
$-$g$ plane for the spin tilting angle $\theta _{{\rm t}}$ (d), the smaller
gap $\Delta _{-}$ (d) and the spin winding number $n_{{\rm w}}^{zx}$ (f).
The dashed lines in panels (d) and (f) are the analytic result from (\ref%
{gR-1}) and (\ref{gR-2}). One sees that there is a $\theta _{{\rm t}}$-jump
across this boundary and the gap is closing (blue belt in panel (e)). One
may also notice that, besides the reversal of $\theta _{{\rm t}}$ in panel
(d), there is also a reversal of the spin winding direction in panel (e) at
the transition boundary. We shall demonstrate in Section \ref{Sect-Winding-Direction-transition} the winding direction reversal more explicitly by the spin winding itself after
extracting other transitions.

\begin{figure*}[t]
\includegraphics[width=2\columnwidth]{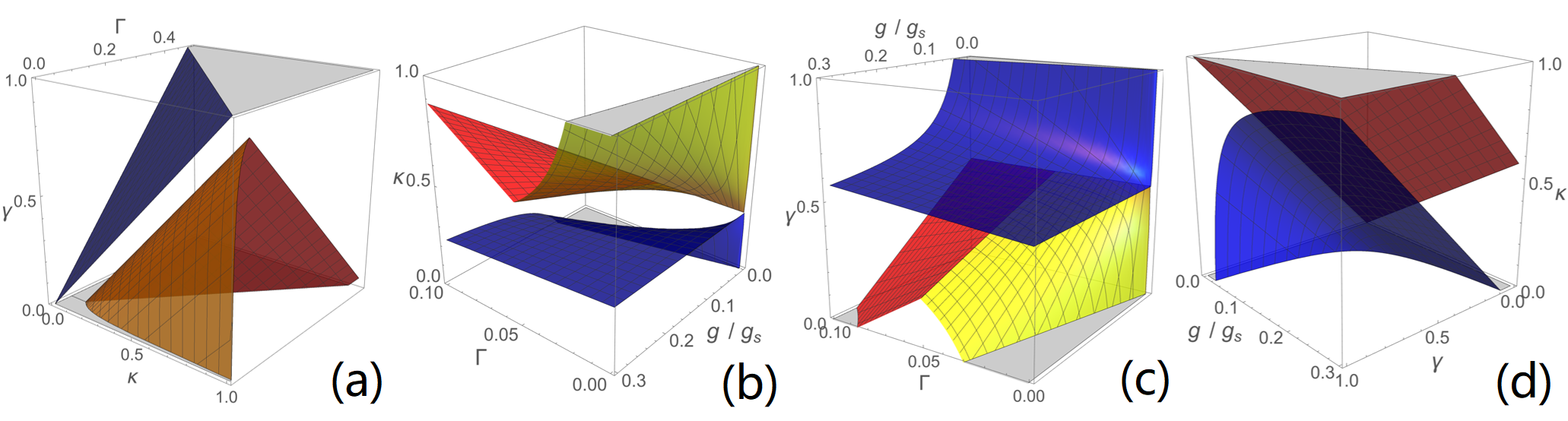}
\caption{3D transition boundaries for the spin winding reversal transition (red), gapped reversal transition (yellow) and super-invariant point (blue) in $\kappa$-$\Gamma$-$\gamma$ (a), $\Gamma$-$g$-$\kappa$ (b), $\Gamma$-$g$-$\gamma$ (c) and $\gamma$-$g$-$\kappa$ (d) spaces. a) $g=0.1g_{\rm s}$, b) $\gamma =0.3$, c) $\kappa =0.5$, d) $\Gamma=0.1$.  In all panels, $\omega=0.9$, $n=1$ and $\eta=-1$, with $\Omega=1$ as the unit. The reversal transition is $n$-dependent while the other two transitions are level-independent.
}
\label{Fig-3D-Diagram}
\end{figure*}

\section{Reversal Transition of Spin Winding without Gap Closing}
\label{Sect-Reversal-No-Gap-Closing}

The reversal transitions in Section \ref{Sect-Reversal-Gap-Closing} occur at
gap closing, in this section we shall reveal another reversal transition
without gap closing. Moreover, this new transition is partially
level-independent in the sense that the transition point is the same for all
the levels that possess this transition (Here by ``partially level-independent" we mean that this transition may vanish after meeting the first reversal transition in Section \ref{Sect-Reversal-Gap-Closing}). Indeed in Figure \ref%
{Fig-Reverse-Transition}a all the lines of tilting angle $\theta _{{\rm t}}$
with different $n$ meet at the same point marked by the dots around $\Gamma
=0.016\Omega $ where there is an angle reversal from $\theta _{{\rm t}}=-\pi
/2$ to $\pi /2$. Note here there is no gap closing as one can see in Figure %
\ref{Fig-Reverse-Transition}b, in contrast to the vanishing gap at the other
reversal transitions.

This gapped reversal (GR) transition appears when $\langle \sigma _{z}\left(
x\right) \rangle $ overall at any position $x$ is going through its zero
point, which can be realized at $\widetilde{C}_{z}=0$ in (\ref{Coefficient-z}%
) with the analytic critical point
\begin{eqnarray}
\kappa _{{\rm GR}} &=&\gamma +g\frac{\left( \Omega -\omega \right) }{\Gamma }%
,\qquad \Gamma _{{\rm GR}}=g\frac{\left( \Omega -\omega \right) }{\left(
\kappa -\gamma \right) },  \label{gSR-1} \\
\gamma _{{\rm GR}} &=&\kappa -g\frac{\left( \Omega -\omega \right) }{\Gamma }%
,\qquad g_{{\rm GR}}=\Gamma \frac{\left( \kappa -\gamma \right) }{\left(
\Omega -\omega \right) }.  \label{gSR-2}
\end{eqnarray}%
In Figure \ref{Fig-Reverse-Transition}d,f, the dotted lines are plotted by
the analytic expressions for the gapped reversal transition in (\ref{gSR-2}), which reproduces the boundary manifested by $\theta _t$ and $n_{\rm w}^{zx}$.
Correspondingly in Figure \ref{Fig-Reverse-Transition}e, there is no gap
closing, in contrast to the gap closing in the blue belt around $g_{{\rm R}}$. Also, besides the
reversal of the tilting angle of the spin winding plane, there is a reversal of spin
winding direction in the $\langle \sigma _{z}\left( x\right) \rangle $-$%
\langle \sigma _{x}\left( x\right) \rangle $ plane as indicated by the sign
changeover of $n_{{\rm w}}^{zx}$ in Figure \ref{Fig-Reverse-Transition}f.

\section{Super-Invariant Point: No Spin Tilting, Level-Independent, and
Without Gap Closing.}

\label{Sect-Super-Invariant}

We find a super-invariant point in the presence of non-Hermiticity, in the
sense that it not only has no tilting of the spin winding plane, as the
Hermitian case, but also is independent of $n$. We give an illustration in
{\bf Figure} \ref{Fig-Super-Invariant}a where all $\theta _{{\rm t}}$ lines
with different $n$ go through a same point with $\theta _{{\rm t}}=0$ in the
variation of $\gamma $, which forms a super-invariant point as marked by the
black dot. The super-invariant point can be seen in Figure \ref%
{Fig-Super-Invariant}b,c in the $\Gamma $-$\gamma $ plane where the
invariant point becomes a line (solid black) which remains the same when we
change the levels ($n=1$ in panel (b) and $n=4$ (c)), Here, by the way, the
dotted lines in (b) and (c) confirm the gapped reversal transition in Section %
\ref{Sect-Reversal-No-Gap-Closing} which is also level-independent, while
the dashed lines denoting the reversal transitions in Section \ref%
{Sect-Reversal-Gap-Closing} are level dependent. There is no spin winding
direction reversal in the $\langle \sigma _{z}\left( x\right) \rangle $-$%
\langle \sigma _{x}\left( x\right) \rangle $ plane, as shown by $n_{{\rm w}%
}^{zx}$ in Figure \ref{Fig-Super-Invariant}e where $n_{{\rm w}}^{zx}$ is
changing the sign only at the $\gamma _{{\rm GR}}$ boundary (dotted) and the
$\gamma _{{\rm R}}$ boundary (dashed) but not at the super-invariant
boundary. Although there is a spin winding direction reversal in the $%
\langle \sigma _{y}\left( x\right) \rangle $-$\langle \sigma _{x}\left(
x\right) \rangle $ plane, the tilting angle is changing slowly rather than
an abrupt jump, as one can see in Figure \ref{Fig-Super-Invariant}a. Here
the seemingly fast change of $\theta _{{\rm t}}$ around the solid lines in
Figure \ref{Fig-Super-Invariant}b,c is due to the plot in amplitude
amplifying by $\left\vert \theta _{{\rm t}}\right\vert ^{1/4}$ to increase
the color contrast and visibility of the super-invariant boundary.

The no-tilting condition can be provided by setting $\widetilde{C}_{y}=0,$
from which we obtain the analytic super-invariant point
\begin{eqnarray}
\kappa _{{\rm SI}} &=&\gamma -\frac{\Gamma \left( \Omega -\omega \right) }{g}%
,\qquad \Gamma _{{\rm SI}}=\frac{g\left( \gamma -\kappa \right) \ }{\left(
\Omega -\omega \right) },  \label{gSI-1} \\
\gamma _{{\rm SI}} &=&\kappa +\frac{\Gamma \left( \Omega -\omega \right) }{g}%
,\qquad g_{{\rm SI}}=\frac{\Gamma \left( \Omega -\omega \right) }{\left(
\gamma -\kappa \right) },  \label{gSI-2}
\end{eqnarray}%
which is plotted in the solid lines and agrees with the overall $\theta _{%
{\rm t}}$ map in Figure \ref{Fig-Super-Invariant}b,c

\section{Reversals of Spin Winding Direction around the Three Special Points}
\label{Sect-Winding-Direction-transition}

So far we have figured out the three special points of the reversal
transition, the gapped reversal transition and the super invariant point.
Although we have mentioned the reversal of spin winding direction for each
point by the sign changeover in the spin winding number $n_{{\rm w}}^{zx}$
and $n_{{\rm w}}^{yx}$, it may provide a more definite and overall view by a
straight demonstration of the spin winding itself for these special points
together. Here in {\bf Figure} \ref{Fig-Winding-direction} the spin windings
in the $\langle \sigma _{z}\left( x\right) \rangle $-$\langle \sigma
_{x}\left( x\right) \rangle $ plane (panels (a-d)) and the $\langle \sigma
_{y}\left( x\right) \rangle $-$\langle \sigma _{x}\left( x\right) \rangle $
plane (panels (e-h)) are plotted for some representative points of the
phases before and after the three special points, as marked by the dots in
Figure \ref{Fig-Super-Invariant}c and labeled by the number $1,2,3,4,$
corresponding to panels (a,e), (b,f), (c,g), (d,g) in Figure \ref%
{Fig-Winding-direction}. The reversal transition lies between points $1$ and
$2$, the gapped reversal transition between $2$ and $3$, and the super
invariant point between $3$ and $4$. Although the spin winding panels
(a,b,d-f,h) has a shape in formal or spiritual similarity with a long or
broad face in long hair and a mask while panels (c,g) look like a face of catgirl,
they are topologically the same except for the spin winding number. Here the
spin winding direction is not only marked by the arrows but also by the
colors with blue for counter-clockwise direction and orange for clockwise
direction. From\ Figure \ref{Fig-Winding-direction}(a-c) we see clearly that
there is a reversal of spin winding direction in the $\langle \sigma
_{z}\left( x\right) \rangle $-$\langle \sigma _{x}\left( x\right) \rangle $
plane both for the reversal transition and the gapped reversal transition,
but from panels (e-g) one finds no direction reversal in the $\langle \sigma
_{y}\left( x\right) \rangle $-$\langle \sigma _{x}\left( x\right) \rangle $
plane; Conversely, from (c-d) there is no spin winding direction reversal
for the super-invariant point in the $\langle \sigma _{z}\left( x\right)
\rangle $-$\langle \sigma _{x}\left( x\right) \rangle $ panels (e-g) but
from (g-h) there is a reversal in the $\langle \sigma _{y}\left( x\right)
\rangle $-$\langle \sigma _{x}\left( x\right) \rangle $ plane.

\section{3D Boundaries}
\label{Sect-3D-Boundaries}

In Figures \ref{Fig-Reverse-Transition} and \ref{Fig-Super-Invariant} only
some 2D boundaries are plotted for the three special points as
afore-discussed, to get a panorama here we give 3D boundaries. In {\bf Figure%
} \ref{Fig-3D-Diagram}, we give some overview boundaries for the spin
winding reversal transition (red), the gapped reversal transition (yellow)
and the super-invariant point (blue) in the $\kappa $-$\Gamma $-$\gamma $, $%
\Gamma $-$g$-$\kappa $, $\Gamma $-$g$-$\gamma $ and $\gamma $-$g$-$\kappa $
dimensions respectively with a fixed $g$, $\gamma $, $\kappa $ and $\Gamma $%
. Basically, spin winding reversal transition and the gapped reversal
transition will meet at some boundaries, as one sees in panels (a-c), while
the super-invariant point is separate from them unless in the special
situation with $\Gamma =0$ as in panels (b,c) where the super-invariant point may meet the
reversal transition at $g=0$ and $\kappa =\gamma $.
Another trend one can see is that the super-invariant point can arise
earlier than the reversal transitions in varying $\kappa $, despite it
appears later in varying $\gamma $. These overview pictures may be helpful in
choosing parameters for manipulation of the three special points.

\section{Conclusions and Discussions}

\label{Sect-Conclusions}

We have studied rigorously the topological feature of the fundamental
JCM in the presence of general non-Hermiticity
which may arise from the dissipation and decay rates of the coupling, the
qubit and the bosonic field. We have found that the topological feature of the
eigenstates are quite robust against the non-Hermiticity. Indeed, the
non-Hermiticity only tilts the spin winding plane, while the spin texture
nodes and the spin winding number, which characterize the topological
feature, remain unaffected in the protection of parity symmetry.

Based on the exact solution of the JCM, we have analytically extracted the
spin texture in the position space which forms the spin winding in the space
of local spin expectation. In comparison with the conventional Hermitian JCM,
of which the spin is winding only in the $\langle \sigma _{z}\left( x\right)
\rangle $-$\langle \sigma _{x}\left( x\right) \rangle $ plane, the presence
of the non-Hermiticity induces an additional winding component in the $%
\langle \sigma _{y}\left( x\right) \rangle $-$\langle \sigma _{x}\left(
x\right) \rangle $ plane. We see that $\langle \sigma _{y}\left( x\right) \rangle
$ is proportional to $\langle \sigma _{z}\left( x\right) \rangle $, which
gives rise to a tilting angle of the total winding plane. The spin texture
components $\langle \sigma _{y}\left( x\right) \rangle $ and $\langle \sigma
_{z}\left( x\right) \rangle $ share the same nodes and the nodes are
invariant in variations of system parameters, which is not only valid in the
Hermitian case but also in the non-Hermitian situation. The spin winding
number is shown to be equal to the excitation number both in the $\langle
\sigma _{z}\left( x\right) \rangle $-$\langle \sigma _{x}\left( x\right)
\rangle $ plane and the additional $\langle \sigma _{y}\left( x\right)
\rangle $-$\langle \sigma _{x}\left( x\right) \rangle $ plane, which endows
the excitation number a topological connotation as the topological quantum
number in both winding planes.

Although the magnitude of spin winding number is unaffected by the
non-Hermiticity, the winding direction may be reversed in non-Hermiticity.
We first find a level-dependent reversal transition of the spin winding. The
non-Hermiticity generates a fractional phase in the square root term of the
eigen wave function, which leads to a jump or reversal of the spin tilting
angle and a reversal of spin winding direction in the $\langle \sigma
_{z}\left( x\right) \rangle $-$\langle \sigma _{x}\left( x\right) \rangle $
plane. The total jump of tilting angle is a fraction of $\pi $, with the
degree of the fraction depending on the energy levels. Such reversal
transition occurs at gap closing, similarly to the conventional topological
phase transitions which require a gap closing as in condensed matter\cite%
{Topo-Wen,Hasan2010-RMP-topo,Yu2010ScienceHall,Chen2019GapClosing,TopCriterion,Top-Guan,TopNori}
and also in light-matter interactions.\cite%
{Ying-2021-AQT,Ying-gapped-top,Ying-Stark-top,Ying-Spin-Winding}

We also find a gapped reversal transition. This transition
point is the same for all the energy levels that possess this transition,
in contrast to the first reversal transition which is level-dependent. This gapped
reversal transition involves a total jump of tilting angle by $\pi $ and is accompanied with
a reversal of spin winding direction in the $\langle \sigma _{z}\left(
x\right) \rangle $-$\langle \sigma _{x}\left( x\right) \rangle $ plane. In
particular, unlike the first reversal transition, this transition occurs in a situation without gap
closing, resembling the unconventional topological phase transitions which
break the traditional condition of gap closing as in some special cases of
condensed matter\cite{Amaricci-2015-no-gap-closing,Xie-QAH-2021} and
light-matter interactions.\cite{Ying-gapped-top,Ying-Stark-top,Ying-Spin-Winding}

Finally we have also revealed a super-invariant point which has no titling angle as the Hermitian case despite in a finite non-Hermiticity.
The super-invariant sense lies in the aspect that it is completely independent of the
energy levels. Such a super-invariant point does not reverse the spin winding
direction in the $\langle \sigma _{z}\left( x\right) \rangle $-$\langle
\sigma _{x}\left( x\right) \rangle $ plane, although it varies the winding
direction in the $\langle \sigma _{y}\left( x\right) \rangle $-$\langle
\sigma _{x}\left( x\right) \rangle $ plane in a slow way instead of an abrupt
jump as in the reversal transitions. The super-invariant point also occurs in
the unconventional situation without gap closing.

The present work has focused on the robustness of the topological feature of all the eigenstates against the non-Hermiticity. In fact the non-Hermiticity
will also influence the transitions in the ground states\cite{Ying-JC-winding} and induce some new effects, which deserves a special discussion in some other work.\cite{Ying-JC-Non-Hermitian-Transition}

As a final discussion it should be mentioned that the model can be implemented in realistic
systems, such as in superconducting circuits and hybrid quantum systems,\cite{you024532,Yimin2018,NonHermitianJCM-2022SciChina}
with access to
ultra-strong couplings possible \cite{Ulstrong-JC-1,Ulstrong-JC-3-Adam-2019,Ulstrong-JC-2} and the dissipation
controllable \cite{NonHermitianJCM-2022SciChina}. The position $x$ can be
represented by the flux of Josephson junctions and the spin texture might be
measured by interference devices and magnetometer.\cite{you024532} With
these possible platforms, the found robust topological feature against the
non-Hermiticity, including the invariant spin winding number and supper
invariant points, may be favorable in practical applications where the
dissipation and decay rates may be unnegligible. The reversal transitions
indicate that one may in turn to utilize the
disadvantageous dissipation and decay rates to reverse the spin winding
direction, which should add a control way for topological manipulation of
light-matter coupling systems. The gapped situation in the gapped reversal transition and
supper invariant point also is favorable in making senors as the gapped
situation could avoid the detrimental time divergent problem in preparing
probe state.\cite{Ying2022-Metrology}

As a closing remark, we speculate that our results may be also relevant for
other systems as the JCM under consideration has the effective
Rashba/Dresselhaus spin-orbit coupling which shares similarity with the
coupling in nanowires~\cite%
{Nagasawa2013Rings,Ying2016Ellipse,Ying2017EllipseSC,Ying2020PRR,Gentile2022NatElec}%
, cold atoms~\cite{Li2012PRL,LinRashbaBECExp2011} and relativistic systems.\cite{Bermudez2007}

\section*{Acknowledgements}

This work was supported by the National Natural Science Foundation of China
(Grants No.~11974151 and No.~12247101).

\appendix

\section{Proof for Reversal of $\theta _{{\rm t}}$ at $\Gamma _{{\rm %
R}}$}
\label{Appendix-Proof-Reversal}

The reversal transition found in Section \ref{Sect-Reversal-Gap-Closing}
occurs due to the $2\pi $ phase gain in the argument $\vartheta $ of the
complex parameter $A-iB$ which leads to a nontrivial $\pi $ phase in the
square root term $S_{r}$.
This $\pi $ phase of $S_{r}$ would result in a jump of the
tilting angle
$\theta _{{\rm t}}$ to some $\theta _{{\rm t}}+\Delta \theta _{{\rm t}}$.
Here we prove this tilting angle jump is actually an angle reversal from
$\theta _{{\rm t}}$ to $-\theta _{{\rm t}}$. We can set $\vartheta
\rightarrow \eta _{{\rm R}}\pi $ with $\eta _{{\rm R}}=\pm 1$
around the reversal transition, e.g. around $\Gamma _{{\rm R}}$ in varying $\Gamma $.
Then
\begin{equation}
\widetilde{C}_{z}\rightarrow \widetilde{C}_{z1}+\eta _{{\rm R}}%
\widetilde{C}_{z2},\quad \widetilde{C}_{y}\rightarrow \widetilde{C}%
_{y1}+\eta _{{\rm R}}\widetilde{C}_{y2}
\end{equation}
where
\begin{eqnarray}
\widetilde{C}_{z1} &=&gd_{\Omega \omega }-\Gamma d_{\kappa \gamma },\quad
\widetilde{C}_{z2}=-2\eta R\Gamma , \\
\widetilde{C}_{y1} &=&\Gamma d_{\Omega \omega }+gd_{\kappa \gamma },\quad
\widetilde{C}_{y2}=2\eta Rg.
\end{eqnarray}
As $\theta _{{\rm t}}=\arctan \frac{\widetilde{C}_{y}}{\widetilde{C}_{z}}$
from (\ref{def-theta-t}), to prove the reversal from $\theta _{{\rm t}}$ to $%
-\theta _{{\rm t}}$ we need to check
\begin{equation}
\frac{\widetilde{C}_{y1}+\widetilde{C}_{y2}}{\widetilde{C}_{z1}+\widetilde{C}
_{z2}}=-\frac{\widetilde{C}_{y1}-\widetilde{C}_{y2}}{\widetilde{C}_{z1}-\widetilde{C}_{z2}}
\end{equation}
which is equivalent to
\begin{equation}
4R^{2}g\Gamma =g\Gamma \left( d_{\Omega \omega }^{2}-d_{\kappa \gamma }^{2}\right) +\left( g^{2}-\Gamma ^{2}\right) d_{\Omega \omega }d_{\kappa
\gamma }.  \label{Proof-Eq-4RgGamma}
\end{equation}
Replacing $\Gamma $ by the reversal point $\Gamma _{{\rm R}}=d_{\Omega
\omega }d_{\kappa \gamma }/\left( 4ng\right) $ from (\ref{gR-1}), Equation (\ref{Proof-Eq-4RgGamma}) becomes
\begin{equation}
\left[ 16R^{2}g^{2}n+\left( 4ng^{2}-d_{\kappa \gamma }^{2}\right) \left(
4ng^{2}+d_{\Omega \omega }^{2}\right) \right] d_{\Omega \omega }d_{\kappa
\gamma }=0.
\end{equation}%
From the reversal transition condition (\ref{R-condition-gmma}) it is easy
to see
\begin{equation}
\left( 4ng^{2}-d_{\kappa \gamma }^{2}\right) \left( 4ng^{2}+d_{\Omega \omega
}^{2}\right) <0.  \label{4ng2-minus}
\end{equation}%
On the other hand, from (\ref{Def-theta})-(\ref{Def-theta}) one can find
\begin{equation}
R^{4}=[n\left( g^{2}-\Gamma ^{2}\right) -\frac{1}{4}d_{\kappa \gamma }^{2}+
\frac{1}{4}d_{\Omega \omega }^{2}]^{2}+[2ng\Gamma -\frac{1}{2}d_{\Omega
\omega }d_{\kappa \gamma }]^{2}
\end{equation}%
which becomes%
\begin{equation}
R^{4}=\frac{\left( 4ng^{2}-d_{\kappa \gamma }^{2}\right) ^{2}\left(
4ng^{2}+d_{\Omega \omega }^{2}\right) ^{2}}{\left( 4g\right) ^{4}n^{2}}
\label{R4@gR}
\end{equation}%
at $\Gamma =\Gamma _{{\rm R}}$ with the result%
\begin{equation}
\left( 16R^{2}g^{2}n\right) ^{2}=\left[ \left( 4ng^{2}-d_{\kappa \gamma
}^{2}\right) \left( 4ng^{2}+d_{\Omega \omega }^{2}\right) \right] ^{2}
\label{R8@gR}
\end{equation}%
With (\ref{4ng2-minus}) and (\ref{R8@gR}) we find
\begin{equation}
\left[ 16R^{2}g^{2}n+\left( 4ng^{2}-d_{\kappa \gamma }^{2}\right) \left(
4ng^{2}+d_{\Omega \omega }^{2}\right) \right] =0,
\end{equation}%
thus completing the proof for the reversal of $\theta _{{\rm t}}$ at the
reversal transition.

\end{document}